\documentclass[prl,showpacs,twocolumn,superscriptaddress]{revtex4-1}
\usepackage[utf8]{inputenc}
\usepackage[american,british]{babel}
\usepackage[T1]{fontenc}
\usepackage[pdftex]{graphicx}  
\usepackage{graphicx, xcolor}
\usepackage{dcolumn}
\usepackage{bm}
\usepackage{amsmath,amsthm,amssymb}
\usepackage{amsmath}
\usepackage{color}
\usepackage{ulem}

\newcommand{\bra}[1]{\left\langle #1 \right|}
\newcommand{\ket}[1]{\left| #1 \right\rangle}

\newcommand{\mean}[1]{\left\langle#1\right\rangle}

\newcommand{\bd}[1]{b^{\dagger}_{#1}}

\newcommand{\ad}[1]{a^{\dagger}_{#1}}

\newcommand{\mytitle}{
Localized Majorana-like modes in a number conserving setting: \\ An exactly solvable model}

\begin{document}

\title{\mytitle}

\author{Fernando Iemini}
\affiliation{Departamento de F\'isica - ICEx - Universidade Federal de Minas Gerais,  Belo Horizonte - MG, Brazil}
\affiliation{NEST, Scuola Normale Superiore \& Istituto Nanoscienze-CNR, I-56126 Pisa, Italy}

\author{Leonardo Mazza}
\affiliation{NEST, Scuola Normale Superiore \& Istituto Nanoscienze-CNR, I-56126 Pisa, Italy}

\author{Davide Rossini}
\affiliation{NEST, Scuola Normale Superiore \& Istituto Nanoscienze-CNR, I-56126 Pisa, Italy}

\author{Rosario Fazio}
\affiliation{NEST, Scuola Normale Superiore \& Istituto Nanoscienze-CNR, I-56126 Pisa, Italy}

\author{Sebastian Diehl}
\affiliation{Institute of Theoretical Physics, TU Dresden, D-01062 Dresden, Germany}

\date{\today}

\begin{abstract} 
In this letter we present, in a number conserving framework, a model of interacting fermions in a two-wire geometry 
supporting non-local zero-energy Majorana-like edge excitations. The model has an exactly solvable 
line, on varying the density of fermions, described by a topologically non-trivial ground state wave-function. Away from the 
exactly solvable line we study the system by means of the numerical density matrix renormalization group. 
We characterize its topological properties through the explicit calculation of a degenerate entanglement spectrum and of the braiding operators which are exponentially localized at the edges. Furthermore, we
establish the presence of a gap in its single particle spectrum while the Hamiltonian is gapless, and compute the 
correlations between the edge modes as well as the superfluid correlations. The topological phase covers 
a sizeable portion of the phase diagram, the solvable line being one of its boundaries.
\end{abstract}

\maketitle

\paragraph*{Introduction ---}

Large part of the enormous attention devoted in the last years to topological superconductors owes to the exotic quasiparticles such as Majorana modes, which localize at their boundaries (edges, vortices, \ldots)~\cite{hasan10,qi11} and play a key role in several robust quantum 
information protocols~\cite{nayak08}. Kitaev's  $p$-wave superconducting quantum wire~\cite{kitaev01} provides a minimal 
setting showcasing all the key aspects of topological states of matter in fermionic systems. The existence of a so-called 
``sweet point'' supporting an exact and easy-to-handle analytical solution puts this model at the heart of our understanding 
of systems supporting Majorana modes. Various implementations in solid state~\cite{lutchyn10,oreg10} and ultracold 
atoms~\cite{Sato_2009, jiang11} via proximity to superconducting or superfluid reservoirs have been proposed, and experimental signatures 
of edge modes were reported~\cite{mourik12}.

Kitaev's model is an effective mean-field model and its Hamiltonian does not commute with the particle number operator. 
Considerable activity has been devoted to understanding models supporting Majorana edge modes in a number-conserving setting~\cite{fidkowski11,sau11,cheng11,kraus13,Ortiz}, as in various experimental platforms  (e.g. solid state \cite{fidkowski11,sau11} or ultracold 
atoms \cite{cheng11,kraus13}) this property is naturally present. It was realised that a simple way to promote particle number conservation to a 
symmetry of the model, while keeping the edge state physics intact, was to consider at least two coupled wires rather than a 
single one~\cite{fidkowski11,sau11,cheng11}. However, since attractive interactions are pivotal to generate superconducting order 
in the canonical ensemble, one usually faces a complex interacting many-body problem. Therefore, approximations such as  
bosonization~\cite{fidkowski11,sau11,cheng11}, or numerical approaches~\cite{kraus13} were invoked.  An exactly solvable model 
of a topological superconductor in a number conserving setting, which would directly complement Kitaev's scenario, is missing (see however~\cite{Ortiz}).

In this letter we present an exactly solvable model of a topological superconductor which supports exotic Majorana-like quasiparticles
at its ends and retains the fermionic number as a well-defined quantum number. The construction of the Hamiltonian with local two-body 
interactions and of its ground state draws inspiration from ideas on dissipative state preparation for ultracold atomic 
fermions~\cite{diehl10,diehl11,bardyn13}, here applied to spinless fermions in a two-wire geometry. 
The solution entails explicit ground state wave-functions, which feature all the main qualitative properties 
highlighted so far in approximate analytical~\cite{cheng11,ruhman15,keselman15} and numerical~\cite{kraus13,keselman15} studies for this class of models, with the advantage of being easy-to-handle. 

In particular, we establish the following key features:
i) The existence of one/two degenerate ground states depending on the periodic/open 
boundaries  with a two-fold degenerate entanglement spectrum; 
ii) the presence of exponentially localized, symmetry-protected 
edge states and braiding matrices associated to this degeneracy; 
iii) exponential decay of the fermionic single particle correlations, even if the Hamiltonian is gapless with collective, quadratically dispersing  bosonic 
modes; iv) $p$-wave superconducting correlations which saturate at large distance. 

\begin{figure}[t]
\centering
\includegraphics[width=\columnwidth]{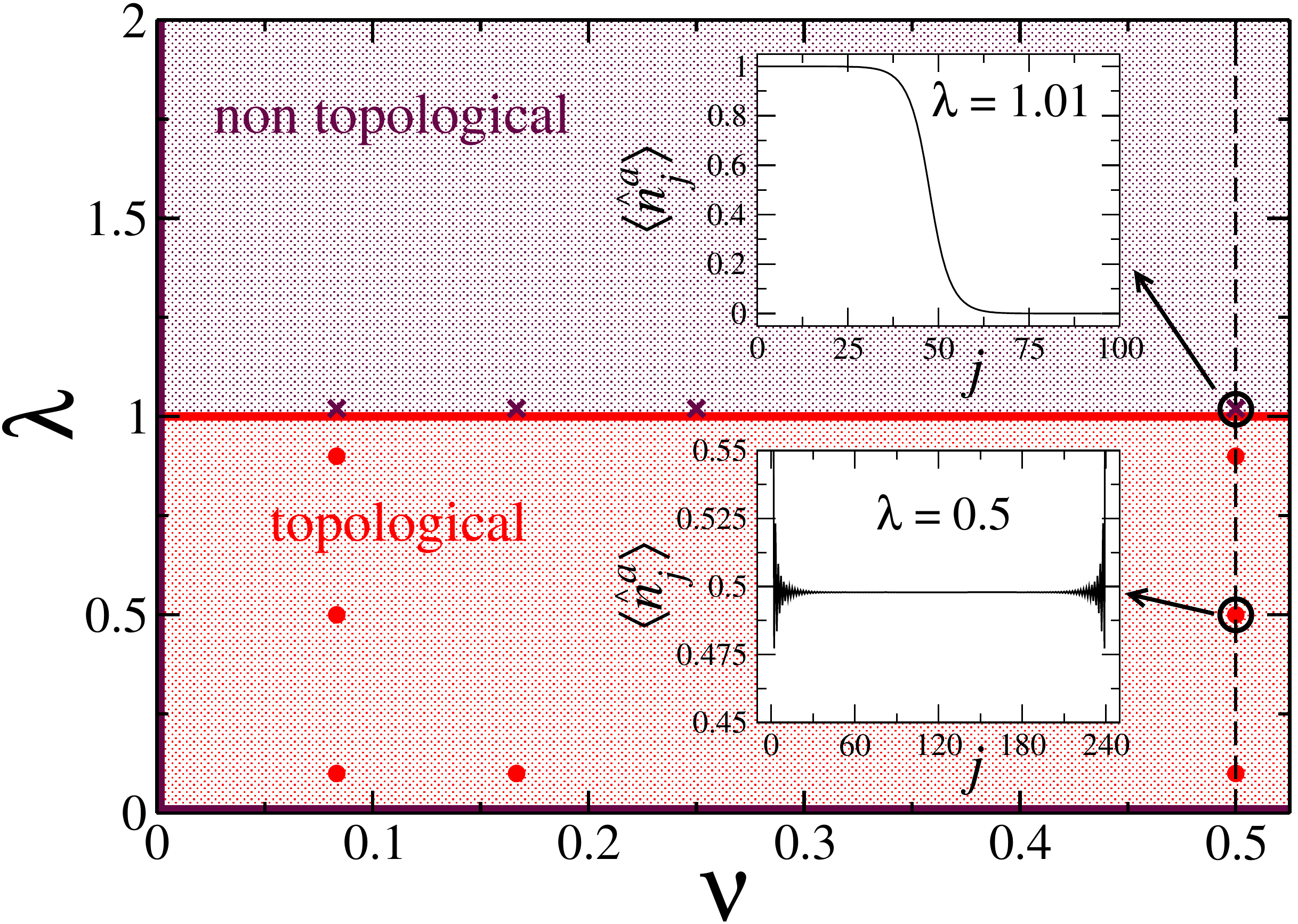}
\caption{Phase diagram for the number and local parity conserving two-wire model~\eqref{eq:hamiltonian} as a function of $\lambda$ and filling $\nu= N/2L$ obtained through DRMG simulations. The exactly solvable topological line is at $\lambda = 1$ (another, trivially solvable non-topological line is at $\lambda=0$). For $\lambda>1$, the system undergoes phase separation (see the density profile $\langle \hat n_j^a \rangle$ in the inset). For $0<\lambda<1$ and $\nu \neq 0$, $1$, the system is in a homogeneous topological phase (see inset). The phase diagram is symmetric with respect to half filling $\nu=1/2$ due to particle-hole symmetry of $\hat H_{\lambda}$.}
\label{fig:phasediagram}
\end{figure}

By tuning the ratio of interaction vs. kinetic energy of our model, we can explore its properties outside the exactly-solvable line. The full phase diagram (Fig.~\ref{fig:phasediagram}) is obtained 
by means of density matrix renormalization group (DMRG) calculations. 
The exactly solvable line is found to stand between a stable topological phase 
and a phase-separated state.

\paragraph*{The model ---} 

We begin by recapitulating some properties of the Kitaev chain, whose Hamiltonian reads~\cite{kitaev01}
\begin{displaymath}
  \hat H_{\rm K} = \sum\limits_{j} \big[ -J \hat a_{j}^\dagger \hat a_{j+1}  -
    \Delta \hat a_{j} \hat a_{j+1} + {\rm H.c.}  - \mu \big(\hat n_j-1/2 \big) \big]. 
\end{displaymath}
Here, $J>0$ denotes the hopping amplitude, $\mu$ and $\Delta$ the chemical potential and the superconducting gap, respectively; 
$\hat a_j^{(\dag)}$ are fermionic annihilation (creation) operators on site $j$, and $\hat n_j\equiv \hat a_{j}^\dagger \hat a_j$. 
This model has i) two density-driven phase transitions from 
finite densities to the empty and full states at $|\mu| = 2J$
(for $\Delta \neq 0$), and ii) a transition driven by the competition of 
kinetic and interaction energy (responsible for pairing) at $\Delta/J=0$
(for $|\mu| < 2J$). 
For $|\mu|<2J$ and $\Delta \neq 0$,
the ground state is unique for periodic boundary conditions, 
but twofold degenerate for open geometry, hosting localized zero-energy Majorana modes. This topological phase is symmetry protected by 
total fermionic parity $\hat P = (-1)^{\hat N}$, where $\hat N \equiv \sum_j \hat n_j$.

Let us focus on the so-called ``sweet point'', namely 
$\mu =0$, and $\Delta = J >0$ and real, which enjoys the property
$\hat H_{\rm K}  = (J/2) \sum_j \hat \ell_j^\dag \hat \ell_j $ with $\hat \ell_j = \hat C_j^\dag + \hat A_j$,  $\hat C_j^\dagger = \hat a^\dag_j + \hat a^\dag_{j+1}$ and  $\hat A_j = \hat a_j - \hat  a_{j+1}$
($\hat \ell_L$ is defined identifying $L+1 \equiv 1$).
For open geometry, the two ground states with $L$ sites 
satisfy $\hat \ell_j |\psi\rangle = 0$, for $1\leq j \leq L-1$, 
and can be written~\cite{turner11} as the equal weighted superposition 
of all even ($e$) or odd ($o$) particle number states: 
\begin{equation}
  \label{eq:wfnoncons}
     \ket{\psi}_{e(o)} = \mathcal{N}_{e(o),L}^{-1/2} \sum_{n}(-1)^{n}
     \sum_{\{\vec{j}_{2n(2n+1)}\}}  \ket{\vec{j}_{2n(2n+1)}} \,.
\end{equation}
Here $|\vec{j}_{m} \rangle = \hat a_{j_1}^\dag \hat a_{j_2}^\dag ... \hat a_{j_m}^\dag \ket{\text{vac}}$ with $j_i<j_{i+1}$ ($j_i = 1, \ldots, L$) and  
$\mathcal{N}_{e,L} = \sum_{n} \binom{L}{2n}$; $\mathcal{N}_{o,L} = \sum_{n} \binom{L}{2n+1}$. 

We now turn to a number conserving version of this model on a single wire~\cite{diehl11}. Indeed, the following model reduces precisely to the above scenario upon performing a naive BCS mean field treatment. Consider the Hamiltonian
$ \hat H_{\rm K}' \equiv \sum_j \hat L_j^\dag \hat L_j$, with $\hat L_j = \hat C_j^\dag \hat A_j $,
whose exact ground state wave-functions can be obtained as follows. Since $\hat A_j |\psi\rangle_{e(o)}  = - \hat C^\dag_j  |\psi\rangle_{e(o)} $, 
$|\psi\rangle_{e(o)} $ are also ground states of $\hat H_{\rm K}'$: $\hat L_i |\psi\rangle_{e(o)}  =0$ because $(\hat C_j^\dag)^2 =0$. 
As $\hat L_i$ conserves the 
particle number, $[\hat L_i , \hat N ]=0$, we can classify ground states for each fixed particle number sector $N$ by number projection, 
$|\Psi,N\rangle = \hat P_N |\psi\rangle_{e(o)}$. This is implemented by choosing the state 
with $2n =N$ (or $2n+1=N$) in the sum over $n$ in Eq.~\eqref{eq:wfnoncons}, and adjusting the normalization to 
$\mathcal{N}_{L,N} = \binom{L}{N}$. The positive semi-definiteness of $\hat H_{\rm K}'$ implies that these states, having zero energy eigenvalue, 
are ground states. However, once $N$ is fixed, the ground state $|\Psi,N\rangle$ is unique, as follows from the Jordan-Wigner mapping to the Heisenberg model~\cite{sachdevbook}. 
The topological twofold degeneracy is lost. 

Guided by the previous analysis, we construct an exactly solvable topological two-wire model with fermionic operators 
$\hat a_j^{(\dagger)}$, $\hat b_j^{(\dagger)}$. 
In addition to those involving each wire
$\hat L_{a(b), j} = \hat C_{a(b), j}^\dag  \hat A_{a(b), j}$, we introduce new operators 
$\hat L_{I, j} = \hat C_{a, j}^\dag \hat A_{b, j} + \hat C_{b, j}^\dag \hat A_{a, j}$. 
The Hamiltonian 
\begin{equation}
\hat H = \sum_{\alpha =a,b,I} \sum_{j=1}^{L-1} \hat L_{\alpha, j}^\dag \hat L_{\alpha, j}
\end{equation}
coincides with the $\lambda = 1$ point of the following more general model:
\begin{align}
  \label{eq:hamiltonian}
  \hat H_\lambda \! = \! &\hspace{-0.3cm} - 4 \hspace{-0.3cm}\!\! \sum_{\substack{j=1 , \alpha =a,b}}^{L-1} \!\! \Big[ (\hat \alpha^\dagger_{j}\hat \alpha_{j+1} \! + \! \text{H.c.}) - \! (\hat n_j^{\alpha} + \hat n_{j+1}^{\alpha}) + \! \lambda \hat n_j^{\alpha} \hat n_{j+1}^{\alpha} \Big] \nonumber \\
  & -2 \lambda \sum_{j=1}^{L-1} \Big[ (\hat n_j^a + \hat n_{j+1}^a)(\hat n_j^b + \hat n_{j+1}^b) - (\hat a^\dagger_{j}\hat a_{j+1} \hat b^\dagger_{j} \hat b_{j+1} \nonumber\\ 
    & \hspace{0.65cm} + \hat a_{j}^\dagger \hat a_{j+1} b^\dagger_{j+1} \hat b_{j} - 2 \hat b^\dagger_{j} \hat b^\dagger_{j+1} \hat a_{j+1} \hat a_{j} + {\rm H.c.}) \Big] .
\end{align} 
$\hat H_\lambda$ conserves the total particle number $\hat N = \hat N_a + \hat N_b$
and the local wire parities $\hat P_{a,b}=(-1)^{\hat N_{a,b}}$, which act as protecting symmetries for the topological phase.
The coupling $\lambda$ tunes the relative strength of the kinetic and interaction terms similarly to $\Delta/J$ in $\hat H_{\rm K}$.
Although only $\lambda=1$ is exactly solvable, we will later consider $\lambda \neq 1$ to explore the robustness of the analytical results. 
The phase diagram  is anticipated in Fig.~\ref{fig:phasediagram}.

\paragraph*{Exact results for $\lambda =1$ ---}

For a fixed particle number $N$ and open boundaries, the ground state of $\hat H$ is twofold degenerate, due to the freedom in choosing the  local parity. For even $N$, the ground states read
\begin{align}
  \label{eq:2wireWF}
  \ket{\psi_L(N)}_{ee} \! &= 
  \mathcal{N}_{ee,L,N}^{-1/2} 
  \sum_{n=0}^{N/2}  \!\!
  \sum_{\substack{\{\vec{j}_{2n}\} , \\\{ \vec{q}_{N-2n}\}}} 
  |\vec{j}_{2n}\rangle_{ a} 
  \otimes \ket{\vec{q}_{N-2n}}_b 
  , \\
  \ket{\psi_L(N)}_{oo} \! &=
  \mathcal{N}_{oo,L,N}^{-1/2}
  \sum_{n=0}^{{N/2-1}}  \!\!
  \sum_{\substack{\{\vec{j}_{2n+1}\} ,\\ \{ \vec{q}_{N-2n-1}\}}}
  |\vec{j}_{2n+1}\rangle_{ a} 
  \otimes \ket{\vec{q}_{N-2n-1}}_b \nonumber
\end{align}
where $\mathcal{N}_{ee,L,N} = \sum_{n=0}^{N/2} \binom{L}{2n} \binom{L}{N-2n}$; 
$\mathcal{N}_{oo,L,N} = \sum_{n=0}^{{N/2-1}} \binom{L}{2n+1} \binom{L}{N-2n-1}$. 
The states $|\vec{j}\rangle_{a}$ and $\ket{\vec{q}}_b$ are simple generalizations of the states
 $|\vec{j}\rangle$ defined in Eq.\eqref{eq:wfnoncons} to the wire a and b respectively.
These describe the cases of 
even ($ee$) or odd ($oo$) particle numbers in each of the wires. For odd $N$, the ground states $\ket{\psi_L(N)}_{eo(oe)}$ with an even (odd) 
number of particles in either wire take the identical sum structure as above with the normalization $\mathcal{N}_{ee,L,N}$ in both cases. 
The wave-functions~\eqref{eq:2wireWF} are the unique ground states of the model~\cite{SM}.
An interesting interpretation of $\ket{\psi_L(N)}_{\sigma\sigma'}$ is in terms of number projection of the ground state of 
two decoupled even-parity Kitaev chains $|\text{G}\rangle = \ket{\psi}^a_{e} \otimes \ket{\psi}^b_{e}$:
\begin{eqnarray}
  \ket{\psi_L(N)}_{ee} & \propto & \hat P_N \ket{\text{G}}; \hspace{0.8cm}
  \ket{\psi_L(N)}_{oo}   \propto   \hat P_N \hat \ell_L^{a\,\dagger} \hat \ell_L^{b\,\dagger} \ket{\text{G}}; \nonumber\\
  \ket{\psi_L(N)}_{oe} & \propto & \hat P_N \hat \ell_L^{a\,\dagger} \ket{\text{G}}; \hspace{0.3cm}
  \ket{\psi_L(N)}_{eo} \propto \hat P_N  \hat \ell_L^{b\,\dagger} \ket{\text{G}};\label{quadratic.two.wires.dark.state}
\end{eqnarray}
where $\hat \ell^a_L$ and $\hat \ell^b_L$ 
are the zero-energy modes of the decoupled Kitaev wires at half filling.
This interpretation provides intuition that the two-fold ground-state degeneracy is absent for periodic boundary conditions: since on a circle $\hat H_{\rm K}$ has a unique ground state, the ground state of $\hat H$ with $N$ particles is also unique~\cite{SM}.

\begin{figure}
\centering
\includegraphics[width=\columnwidth]{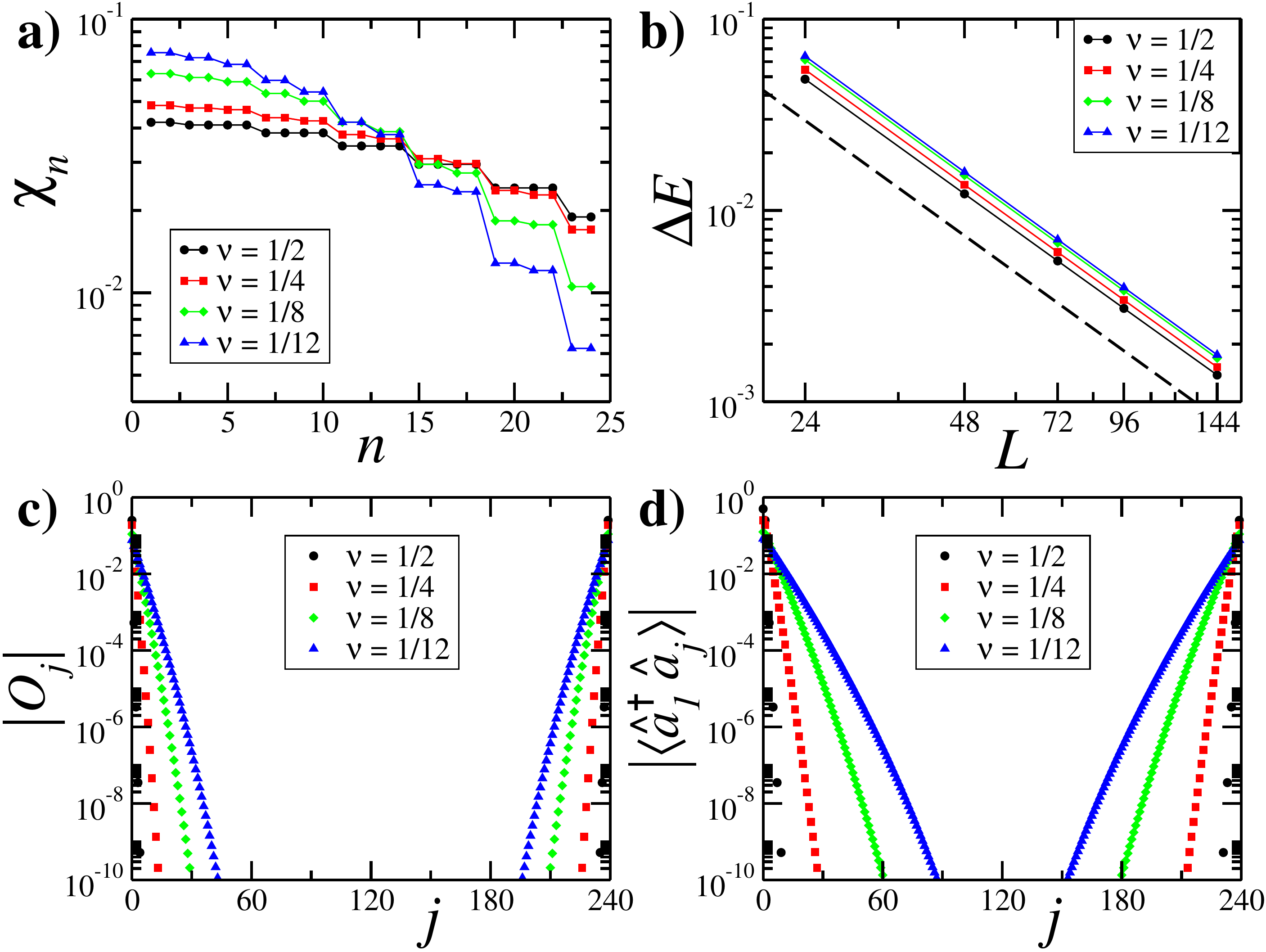}
\caption{
Analysis of model $\hat H$. 
(a) Entanglement spectrum for a reduced state $\rho_l$ with $l = {60}$ for $L=240$.
(b) DMRG results for the scaling of the gap computed at fixed parity, which is compatible with $L^{-2}$ (dashed line); here the number of kept states is $m=400$.
(c) Localization of the edge mode computed via $|\bra{\psi_{L}(N)}_{oo} \hat a^\dagger_j \hat b_j \ket{\psi_{L}(N)}_{ee}|$.
(d) Single-fermion edge correlations $|\langle \hat a^\dagger_1 \hat a_{j} \rangle|$ computed for a system of size $L=240$. The wave-function is shift invariant, such that $|\langle \hat a^\dagger_i \hat a_{i+j} \rangle| \equiv |\langle \hat a^\dagger_1 \hat a_{j} \rangle|$ ($i+j \leq L$).   } 
\label{fig:exact}
\end{figure}

Important evidence of a topologically nontrivial bulk state is obtained from the double degeneracy of the entanglement spectrum, which we now compute for one of the wave-functions~\eqref{eq:2wireWF}.
To this end, we consider the reduced state of $l$ sites on each wire $\rho_l = \text{Tr}_{(L-l)}\big[\ket{\psi_{L}(N)}_{ee} \bra{\psi_{L}(N)}_{ee}\big]$. Taking the symmetries into account, it can be written in diagonal form as~\cite{SM}
\begin{equation}
\rho_l = \sum\limits_{N_l=0}^{min(2l,N)} \sum_{\sigma,\sigma'} \, \chi_{(\sigma \sigma',l,N_l)}^{(L,N)} \ket{\psi_l(N_l)}_{\sigma \sigma'} \bra{\psi_l(N_l)}_{\sigma \sigma'} \label{eq:state:es}
\end{equation}
with the following nonzero eigenvalues:
for $N_l$ even $\chi_{(ee(oo),l,N_l)}^{(L,N)} = \mathcal{N}_{ee(oo),l,N_l} \mathcal{N}_{ee(oo),L-l,N-N_l}/\mathcal{N}_{ee,L,N}$
whereas for $N_l$ odd $\chi_{(eo,l,N_l)}^{(L,N)} = \chi_{(oe,l,N_l)}^{(L,N)} = \chi_{(ee,l,N_l)}^{(L,N)}$.
In the odd-particle number sector the entanglement spectrum is manifestly twofold degenerate. In the even one, such degeneracy appears in the thermodynamic limit:
$\chi_{(ee,l,N_l)}^{(L,N)} / \chi_{(oo,l,N_l)}^{(L,N)} \rightarrow 1$ (see~\cite{SM} and Fig.~\ref{fig:exact}a).

An interesting insight is provided by
$O_j \equiv
\bra{\psi_L(N)}_{oo} \hat a_j^\dag \hat b_j \ket{\psi_L(N)}_{ee}$,
where $\hat a_j^\dag \hat b_j$ is the only single-site operator
which commutes with $\hat N$ and changes the local parities $\hat P_{a,b}$, so that the two ground states can be locally distinguished.
The calculation of such matrix elements leads to a lengthy combinatorial expression~\cite{SM}
and is shown in Fig.~\ref{fig:exact}c. 
We interpret the exponential decay of $O_j$ into the bulk 
as a clear signature of localized edge modes with support in this region only. At half filling the edge states are maximally localized, but away from half filling the number projection increases the localization length. 
In the thermodynamic limit, this length diverges for $\nu \equiv N/2L\to 0,1$, indicating a topological phase transition. 
We emphasize that this exponential behavior is different from~\cite{fidkowski11,sau11}, reporting algebraic localization of the edge states, but similar 
to~\cite{kraus13,keselman15}.
Non-local correlations of edge states are another clear indication of topological order and can be proven via $\langle \hat a^\dagger_1 \hat a_j \rangle$, which is sizeable both at  $j \sim 1$ and $j \sim L$ (see the analytical expression in~\cite{SM} and Fig.~\ref{fig:exact}d).

Furthermore, the Hamiltonian is gapless and hosts long wavelength 
collective bosonic excitations, while the single fermion excitations 
experience a finite gap. This is a crucial property of 
the ground state; the absence of gapless fermion modes in 
the bulk ensures the robustness 
of the zero energy edge modes, in analogy to non-interacting 
topologically non-trivial systems. The gapped nature of single 
fermion excitations is established via the exponential decay of the 
fermionic two-point function, e.g. $\langle \hat a^\dagger_{i} \hat a_{j}\rangle$. Again, the resulting formula is a lengthy combinatorial expression~\cite{SM}, evaluated numerically for very large sizes and plotted in Fig.~\ref{fig:exact}d. For $\nu \to 0,1$, the correlation length diverges, indicating the vanishing of the fermion gap and a thermodynamic, density-driven phase transition in full analogy to the Kitaev chain.

\begin{figure}[!t]
\includegraphics[width=\columnwidth]{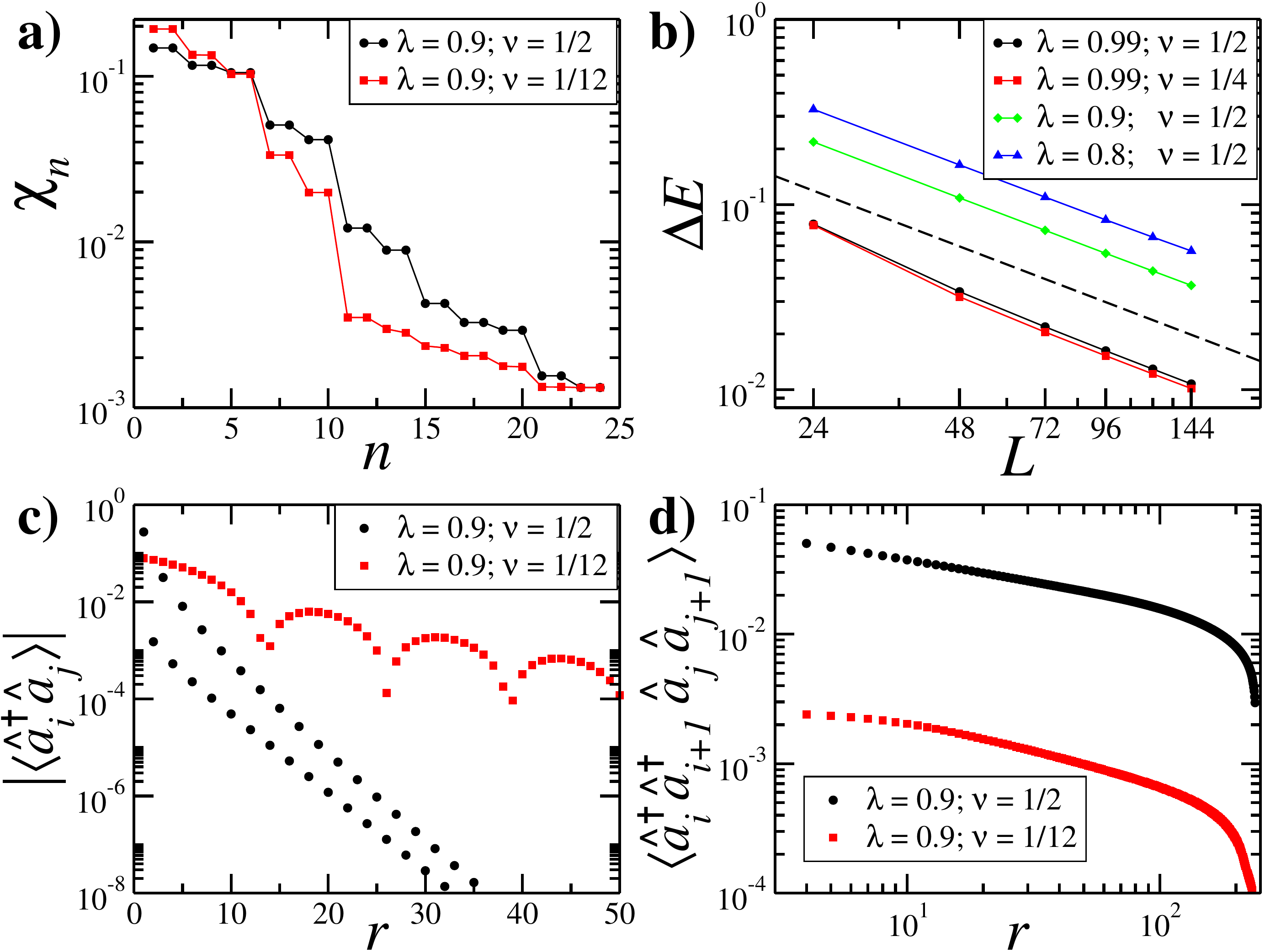}
\caption{DMRG results for model $\hat H_{\lambda}$.
(a) Entanglement spectrum for a reduced state $\rho_l$ with $l = 100$ for $L=240$ ($m=300$).
(b) Algebraic scaling of the gap computed at fixed parity, which is compatible with $L^{-1}$ (dashed line). Here $m=420$.
(c, d) Single-particle $|\langle \hat a^\dagger_i \hat a_j\rangle|$ and p-wave
superfluid $\langle \hat a^\dagger_i\hat a^\dagger_{i+1} \hat a_j\hat a_{j+1}\rangle$ correlations 
at distance $r = |i-j|$ computed in the bulk of system with $L=240$ ($m=300$).
Analogous data were obtained for other values of $\nu$, $\lambda$ (red circles in Fig.~\ref{fig:phasediagram}).
}
\label{fig:DMRG}
\end{figure}

On the other hand, the analysis of the superfluid correlations demonstrates the existence of gapless modes. 
The p-wave nature of these correlations follows from the correlation of the pairing operator $\hat a_{j+1} \hat a_j$. A direct calculation~\cite{SM} shows a saturation at large distance
%
\begin{equation}
  \mean{\hat a^\dagger_{i} \hat a^\dagger_{i+1} \hat a_{j+1} \hat a_j}  \overset{L\rightarrow \infty}{\longrightarrow}  \nu^2(1-\nu)^2 \; .
  \label{eq:Superfluid}
\end{equation}
Similar expressions hold for cross-correlations between the wires. The finite asymptotic value in Eq.~\eqref{eq:Superfluid} hints at the absence of 
bosonic modes with linear dispersion, which would lead to algebraic decay. 
A DMRG analysis of the excitation spectrum of $\hat H$ for system sizes up to $L=144$ 
demonstrates a vanishing of its gap $\sim L^{-2}$ (Fig.~\ref{fig:exact}b). 
This indicates the presence of collective excitations with quadratic dispersion. 
Further support to this statement is provided 
from the fact that Eq.~\eqref{eq:hamiltonian} without the wire coupling term reduces to the XXZ model at the border of its 
ferromagnetic phase, which hosts quadratically dispersing spin waves, $\omega \sim q^2$. This dispersion, with dynamic exponent 
$z=2$, gives rise to an effective phase space dimension $d_\text{eff} = z + 1=3$ at zero temperature, explaining the constancy of 
superfluid correlations due to the absence of a divergence in the soft mode correlators. 
This finding is special for $\lambda=1$. 

\paragraph*{Non-abelian statistics} ---
We now proceed to demonstrate that the edge modes obey a non-abelian statistics completely equivalent to that of Majorana fermions -- i.e., Ising anyons.
Consider the operator 
 $\hat{B}_{aR,bR}(j) = (\hat{\mathcal{I}} + \hat{Z}_{aR,bR,j})/\sqrt{2}$ with $j < L/2$, 
 where  $\hat{Z}_{aR,bR,j} = (\sum_{p=1}^{j} [ \prod_{q=0}^{p-1} \, \hat{Y}_{aR,bR,q} ]
 \, \hat{X}_{aR,bR,p})/\mathcal{F}(j)$, with  
 $\hat X_{aR,bR,j} = 
  ( \ad{L+1-j}b_{L+1-j} - \bd{L+1-j}a_{L+1-j} )$,  
  $\hat Y_{aR,bR,j} 
   = n_{L+1-j}^a n_{L+1-j}^b + (1-n_{L+1-j}^a)(1-n_{L+1-j}^b)$ for $j>0$, 
   $\hat Y_{aR,bR,0} = \hat {\mathcal I}$
 and $\mathcal{F}_j =  \sqrt{1 - [\nu^{2j} + (1-\nu)^{2j}]}$.
$\hat{B}_{aR,bR}(j)$ is thus exponentially localized at the right edge of the ladder and
an analogous operator $\hat{B}_{aL,bL}(j)$ can be defined at the left edge through the transformation mapping an operator at site $l $ to site $L+1-l$ (and viceversa).
Similarly, the operators $\hat{B}_{aR,aL}(j)$ and $\hat{B}_{bL,bR}(j)$ can be defined through the transformations $b_{L+1-l} \to - i a_{l}$ and  $a_{L+1-l} \to - i b_{l}$, respectively.
In general, one can define operators $\hat{B}_{m \Lambda ,m' \Lambda'}(j)$ with $m,m'=a,b$ and $\Lambda, \Lambda' = L,R$.
These operators have the following key properties: They (i) are
exponentially localized at the edges, (ii) act unitarily in 
the ground state subspace, (iii) are
particle number conserving, and (iv) most importantly, provide a 
representation of Majorana braiding operators. 
From this we conclude that the localized edge modes behave as non-abelian Majorana fermions~\cite{Kitaev06}, in full analogy to the case of two neighboring Kitaev wires.
Properties (i)--(iii) 
are demonstrated in~\cite{SM}, whereas here we focus on (iv).
Strictly speaking, properties (ii) and (iv) are only true apart from an error which is exponentially small in $j$ and $L$, which can always be made negligible. In this case we can define the braiding operator $\hat R_{m\Lambda,m'\Lambda'} \equiv \hat{B}_{m\Lambda,m'\Lambda'}(j)$.
We initialize the system in the state $\ket{\psi_L(N)}_{ee}$ and then 
perform two braiding operations on the edges in different sequences. If we consider for example $\hat R_{aR,aL}$ and $\hat R_{aR,bR}$
we obtain $[\hat R_{aR,aL}, \hat R_{aR,bR}] \ket{\psi_L(N)}_{ee} = i \ket{ \psi_L(N)}_{oo}$~\cite{SM}
which demonstrates the non-abelian character 
of these operations. Moreover, this is the pattern 
that the conventional braiding operators produce on two neighboring Kitaev wires
 $\hat R'_{m\Lambda,m'\Lambda'} = e^{\frac{\pi}{4} \gamma_{m\Lambda}\gamma_{m'\Lambda'}} = (\mathcal I + \gamma_{m\Lambda}\gamma_{m'\Lambda'})/\sqrt 2$, where $\gamma_{m\Lambda}$ are Majorana 
 operators fulfilling the Clifford algebra appearing at the left and right ($\Lambda = L,R$) edges of two Kitaev wires $m=a,b$. 
 This pattern coincides for the application of $[\hat R_{m\Lambda,m'\Lambda'}, \hat R_{n \Upsilon,n' \Upsilon'}] $ on all $\ket{\psi_L(N)}_{\sigma \sigma'}$ states (see e.g.~\cite{laflamme14}). In other words, the operators $\hat R_{m\Lambda,m'\Lambda'}$ form a 
 number-conserving representation of Majorana braiding operators on the ground state subspace.

\paragraph*{Numerical results ---}

To further explore the status of these results, we now move to the full model $\hat H_{\lambda}$ away from the solvable line $\lambda = 1$.
The study is performed with DMRG on systems with sizes up to $L=240$ and open boundary conditions.

We first establish the absence of a topological phase for $\lambda > 1$. The density profile, shown in the inset of Fig.~\ref{fig:phasediagram} for $\nu = 0.5$ and $\lambda = 1.01$, displays a clear phase-separation tendency. Analogous data are obtained for other values of $\nu$ (see dark crosses in Fig.~\ref{fig:phasediagram}).
These results can be intuitively understood considering that
$\hat H_{\lambda>1}$ without interwire coupling can be mapped to a gapped ferromagnetic XXZ model with domain walls dual to fermionic phase separation.

For $\lambda < 1$, simulations support the existence of a homogeneous phase (Fig.~\ref{fig:phasediagram}). Note that $\lambda=0$ is a free-fermion point trivially non-topological. For $\lambda \neq 0$ we observe: 
i) two quasi-degenerate ground states with different relative parity and same particle numbers, 
ii) degenerate entanglement spectrum, 
iii) a gap closing as $L^{-1}$ for fixed parity,
iv) exponentially decaying single-fermion correlations, 
v) power-law decaying superfluid correlators.
Plots in Fig.~\ref{fig:DMRG} display our numerical results. 
Simulations at lower filling $\nu \to 0$ and small $\lambda$ are more demanding, owing to the increasing correlation length of the system. 
The numerics is consistent with the phase diagram in Fig.~\ref{fig:phasediagram} exhibiting a topological phase delimited by three trivial lines at $\lambda = 0$, $\nu = 0$ and $\nu = 1$ and an inhomogeneous non-topological phase for $\lambda > 1$. 
The exactly solvable topological line at $\lambda =1$ serves as a boundary; the fixed-$\nu$ phase diagram is reminiscent of the ferromagnetic transition in the XXZ model.

\paragraph*{Conclusions ---} 

We presented an exactly solvable two-wire fermionic model which conserves the number of particles and features Majorana-like exotic quasiparticles at the edges.
Our results can be a valuable guideline to understand topological edge states in number conserving systems.
For example,  the replacement $\hat a_i \to \hat c_{i,\uparrow}, \hat b_i \to \hat c_{i,\downarrow}$ results in a one-dimensional spinful Hubbard Hamiltonian without continuous spin rotation, but time reversal symmetry. The resulting model with an exactly solvable line belongs to the class of time reversal invariant topological superconductors~\cite{qi09}, analyzed in a number conserving setting recently~\cite{keselman15}, with edge modes protected by the latter symmetry. Moreover, exactly solvable number conserving models can be constructed in arbitrary dimension. 

During the final step of preparation, we became aware of similar results obtained by Lang and B\"uchler~\cite{Buchler}.

\paragraph*{Acknowledgments ---} 
We thank E. Altman, M. Baranov, E. Berg, J. C. Budich, M. Burrello, M. Dalmonte M Herrmanns, and G. Ortiz for enlightening discussions. F.I. acknowledges financial support by the 
Brazilian agencies FAPEMIG, CNPq, and INCT- IQ  (National Institute of Science and Technology for Quantum Information). S.D. acknowledges 
support via the START Grant No. Y 581-N16 and the German Research Foundation through ZUK 64. R.F. acknowledges financial support from 
the EU integrated project SIQS and from Italian MIUR via PRIN Project 2010LLKJBX. D.R. and L.M. acknowledge the Italian MIUR through 
FIRB project RBFR12NLNA.
L.M. was supported by Regione Toscana
POR FSE 2007-2013.

\clearpage
\newpage

\clearpage 
\setcounter{equation}{0}%
\setcounter{figure}{0}%
\setcounter{table}{0}%
\renewcommand{\thetable}{S\arabic{table}}
\renewcommand{\theequation}{S\arabic{equation}}
\renewcommand{\thefigure}{S\arabic{figure}}

\onecolumngrid

\begin{center}
{\Large Supplemental Material to \\ \mytitle }

\vspace*{0.5cm}

Fernando Iemini, Leonardo Mazza, Davide Rossini, Rosario Fazio, Sebastian Diehl

\vspace{1cm}

\end{center}

In this Supplemental Material we provide additional information about some details of the analytical results for the exactly solvable two-wire topological system which have been omitted from the main text.

\section{Two-wire ground state}

\subsection{Hard-wall boundary conditions}

In this section we show that the wave-functions $\ket{\psi_L(N)}_{ee(oo)}$ in Eq.~\eqref{eq:2wireWF} of the main text are the only
ground states of the two-wire Hamiltonian $\hat H_{\lambda=1}$ in the presence of hard-wall boundary conditions. Our proof actively constructs
all of the zero-energy eigenstates of the Hamiltonian,
which are the lowest-energy states because $\hat H_{\lambda=1}\geq 0$. Such states are obtained projecting the grand-canonical ground state of two decoupled Kitaev chains onto a given particle-number sector.

Let us first consider only the operators $\{\hat L_{a,j},\hat L_{b,j}\}$ and the corresponding parent Hamiltonian $\hat H_{ab} = \sum_{\alpha = a,b}\sum_j \hat L^{\dagger}_{\alpha, j} \hat L_{\alpha,j}$ which corresponds
to two decoupled wires. We know that the ground states of each wire are given by
$\hat P_N^{\alpha} \ket{\psi}_{\sigma}^\alpha$. 
Hamiltonian $\hat H_{ab}$ thus has a ground space spanned by
\begin{equation}
\left\{\hat P_n^a \hat P_{N-n}^b \ket{\psi}_{\sigma}^a \otimes \ket{\psi}_{\sigma'}^b \right\}_{n=0 }^N
\end{equation}
and $(\sigma, \sigma')$ take the values $(e,e)$ and $(o,o)$ when $N$ is even and $(e,o)$ and $(o,e)$ when $N$ is odd.

An important relation holds because
$\hat l_j^a \ket{\psi}_{\sigma}^a \otimes \ket{\psi}_{\sigma'}^b = 0 $. Upon the insertion of the identity operator $\sum_{n,n'=0}^L \hat P_n^a \hat P_{n'}^b$ we get
\begin{equation}
\sum_{n,n'}^L (\hat C_{a,j}^\dagger \hat P_{n-1}^a + \hat A_{a,j} \hat P_{n+1}^a)
\hat P_{n'}^b \ket{\psi}_{\sigma}^a \otimes \ket{\psi}_{\sigma'}^b =0.
\end{equation}
Each of the elements in the above sum must vanish independently because of orthogonality, and the important relation
\begin{equation}
\hat C_{a,j}^\dagger \hat P_{n-1}^a \hat P_{n'}^b \ket{\psi}_{\sigma}^a \otimes \ket{\psi}_{\sigma'}^b = 
-\hat A_{a,j} \hat P_{n+1}^a \hat P_{n'}^b \ket{\psi}_{\sigma}^a \otimes \ket{\psi}_{\sigma'}^b,\quad \forall n,n'
\end{equation}
is derived
(the same holds for the wire $b$).

Let us now compute the most generic $N$-fermions state such that $\hat H \ket{\phi_N} = 0$. In general, $\ket{\phi_N}$ must be in the kernel of $\hat H_{ab}$:
\begin{equation}
\ket{\phi_N} = \sum_{n=0}^N \sum_{\sigma,\sigma'} x_{n,\sigma,\sigma'} \hat P_n^a \hat P_{N-n}^b  \ket{\psi}_{\sigma}^a \otimes \ket{\psi}_{\sigma'}^b; \quad
\sum_{n,\sigma,\sigma'} |x_{n,\sigma,\sigma'}|^2=1.
\end{equation}
Imposing now that $\hat H_I \ket{\phi_N} = \sum_j \hat L_{I,j}^\dagger \hat L_{I,j} \ket{\phi_N} = 0$, 
we obtain that:
\begin{eqnarray}
\hat L_{I,j} \ket{\phi_N} &=& (\hat C_{a,j}^\dagger \hat A_{b,j} +
 \hat C_{b,j}^\dagger \hat A_{a,j}) \sum_{n=0}^N \sum_{\sigma,\sigma'}  x_{n,\sigma,\sigma'} \hat P_n^a
\hat P_{N-n}^b \ket{\psi}_{\sigma}^a \otimes \ket{\psi}_{\sigma'}^b, \nonumber\\
&=& \hat C_{a,j}^\dagger \hat A_{b,j} \ket{\phi_N} - 
\hat C_{a,j}^\dagger \hat A_{b,j} \sum_n \sum_{\sigma,\sigma'}  x_{n,\sigma,\sigma'} \hat P_{n-2}^a \hat P_{N-n+2}^b
 \ket{\psi}_{\sigma}^a \otimes \ket{\psi}_{\sigma'}^b, \nonumber \\
&=& \hat C_{a,j}^\dagger \hat A_{b,j} \sum_n \sum_{\sigma,\sigma'}  (x_{n,\sigma,\sigma'} - x_{n+2,\sigma,\sigma'})
\hat P_{n}^a \hat P_{N-n}^b \ket{\psi}_{\sigma}^a \otimes \ket{\psi}_{\sigma'}^b = 0 \iff x_{n,\sigma,\sigma'} =  x_{n+2,\sigma,\sigma'},
\end{eqnarray}
since $\hat C_{a,j}^\dagger \hat A_{b,j} \, \hat P_{n}^a \hat P_{N-n}^b \ket{\psi}_{\sigma}^a \otimes \ket{\psi}_{\sigma'}^b \neq 0$. 
Such a relation uniquely defines any ground state for any fixed local parity 
(even-even, odd-odd, even-odd, odd-even), and thus, for each fixed particle number $N$, there is a double degeneracy related to distinct wire parities.
For example, a general ground state for $2N$ particles is given by
\begin{equation}
  \ket{\phi_{2N}} \propto  \sum_{n} \left[ w_0
    \hat P_{2n}^a \hat P_{2(N-n)}^b \ket{\psi}_{e}^a \otimes \ket{\psi}_{e}^b 
    + w_1 \hat P_{2n+1}^a \hat P_{2N - (2n+1)}^b \ket{\psi}_{o}^a \otimes \ket{\psi}_{o}^b \right],
\end{equation}
and is parametrized by the complex coefficients $w_0$ and $w_1$. 
This writing states explicitly that $\ket{\phi_{2N}}$ of the states $\ket{\psi_L(N)}_{ee}$ and $\ket{\psi_L(N)}_{oo}$ presented in the main text.

An alternative viewpoint on the states $\ket{\phi_N}$ stems from considering two decoupled Kitaev chains at half filling $\mu=0$ and $\Delta=J$. After defining the state $\ket{G} = \ket{\psi}_{e}^a \otimes \ket{\psi}_{e}^b$, where $\ket{\psi}_{e}^\alpha$ is the even-parity ground state of the wire $\alpha$, the four ground states of the model read:
\begin{equation}
\{ \ket{G}, \hat \ell^{a\, \dagger}_L \ket{G}, \hat \ell^{b\, \dagger}_L \ket{G},  \hat \ell^{a \,\dagger}_L \hat \ell^{b\,\dagger}_L \ket{G}\},
\end{equation}
and are related to the edge Majorana fermions $\hat \ell^{\alpha}_L= \hat \gamma_{2L}^{\alpha} + i \hat \gamma_{1}^{\alpha}$, 
where $\{\hat \gamma_{i}^{\alpha}, \hat \gamma_{j}^{\beta}\} = 2 \delta_{ij}\delta_{\alpha\beta}$,  
$\hat \gamma_{2j-1}^{\alpha} = i(\hat \alpha_j - \hat \alpha^{\dagger}_j )$, and $\hat \gamma_{2j}^{\alpha} = \hat \alpha_j + \hat \alpha^{\dagger}_j$, for $\alpha = a,b$.
It is possible to explicitly verify that the zero-energy subspaces of the number-conserving Hamiltonian $\hat {H}_{\lambda=1}$ are given by 
\begin{eqnarray}
  \ket{\phi_N} & \in & \text{span}\left\{\hat P_N \ket{G}, \hat P_N \hat \ell^{a\, \dagger}_L \hat \ell^{b\, \dagger}_L \ket{G}\right\} \quad \text{ for } N \text{ even},\nonumber \\
  \ket{\phi_N} & \in & \text{span}\left\{\hat P_N \hat \ell^{a\, \dagger}_L \ket{G}, \hat P_N  \hat \ell^{b\, \dagger}_L \ket{G}\right\} \quad \text{ for } N \text{ odd}.\nonumber
\end{eqnarray} 

\subsection{Periodic boundary conditions}

The situation changes dramatically when one considers periodic boundary conditions. Let us consider for simplicity the case of an even number of sites.
Then, for even values of $N$ there is one zero-energy state, which is a ground state of the Hamiltonian. For odd values of $N$, no zero-energy state appears even in the thermodynamic limit; this is consistent with the fact discussed in the main text that the single-particle correlators decay exponentially, signaling the presence of a thermodynamic gap for the addition of a single particle.

It is a remarkable and non-generic property of this model that the ground state for $N$ even can be formally written in the same way as $\ket{\psi_L(N)}_{oo}$ in Eq.~(4) of the main text.
The demonstration is based on the fact that $\ket{\psi}_o$ (see Eq.~(1) in the main text) is the unique ground state of a Kitaev chain with periodic boundary conditions ($L$ even) and follows the lines of the demonstration for hard-wall boundaries. Again, it is interesting to observe that:
\begin{equation}
 \ket{\psi_L(N)}_{oo} \propto \hat P_N \ket{\psi}_o^a \otimes \ket{\psi}_o^b
\end{equation}
even with periodic boundary conditions.
The absence of an even-parity ground state for the periodic Kitaev wire provides intuition of why the odd $N$ sector of the Hilbert space is gapped.

\section{Entanglement spectrum}

In this section we provide the detailed derivation for the entanglement spectrum of $\ket{\psi_{L}(N)}_{ee}$ presented in the main text. We consider the reduced state of $l$ sites on each wire $\rho_l = \text{Tr}_{(L-l)}\big[\ket{\psi_{L}(N)}_{ee} \bra{\psi_{L}(N)}_{ee}\big]$ (in the following expression identity operators on the first $l$ sites are omitted):
\begin{equation}\label{red.state.bracket.two.wires}
  \rho_l = \hspace{-0.4cm} \sum_{\{\vec{j}_m\},\{\vec{q}_m'\}} \hspace{-0.3cm} \langle\vec{j}_{m}|_a \otimes \bra{\vec{q}_{m'}}_b  \ket{\psi_{L}(N)}_{ee} 
  \bra{\psi_{L}(N)}_{ee}\, 
|\vec{j}_{m}\rangle_a \otimes \ket{\vec{q}_{m'}}_b,
\end{equation}
where $\{\vec{j}_m\}$ and $\{\vec{q}_m'\}$ represent all possible fermionic configurations in $L-\ell$ sites. Let us now explicitly compute
$\langle\vec{j}_{m}|_a \otimes \bra{\vec{q}_{m'}}_b  \ket{\psi_{L}(N)}_{ee}$; for
even values of $N_{\ell} \equiv N-m-m'$:
\begin{equation}
  \langle\vec{j}_{2m(2m+1)}|_a \otimes \langle\vec{q}_{N-N_l-2m(2m+1)}|_b \,\ket{\psi_{L}(N)}_{ee} = \sqrt{\frac{\mathcal{N}_{ee(oo),l,N_l}}{\mathcal{N}_{ee,L,N}}} \,\ket{\psi_l(N_l)}_{ee(oo)};
  \label{eq:ES:even}
\end{equation}
which remarkably does not depend on the specific $\vec j$ or $\vec q$. A similar relation exists for $N_l$ odd,
\begin{equation}
  \langle\vec{j}_{2m(2m+1)}|_a \otimes \langle\vec{q}_{N-N_l-2m(2m+1)}|_b \,\ket{\psi_{L}(N)}_{ee} = \sqrt{\frac{\mathcal{N}_{ee,l,N_l}}{\mathcal{N}_{ee,L,N}}} \,\ket{\psi_l(N_l)}_{eo(oe)},
  \label{eq:ES:odd}
\end{equation}
Summing up such terms, we obtain the reduced state in diagonal form and its eigenvalues, presented in the main text.

The demonstration for the double degeneracy in the entanglement spectrum in the limit of large $L$ and $L-l$ (i.e., large lattices and bipartitions not close to its edges), is related to the fact that, in this limit, $\mathcal{N}_{ee,l,N_l} \sim \mathcal{N}_{oo,l,N_l}$.
Even if we do not have an explicit analytical proof of the previous relation, numerical tests in several regimes corroborate this intuitive result.

\begin{figure}[t]
\begin{center}
\includegraphics[width=0.5\columnwidth]{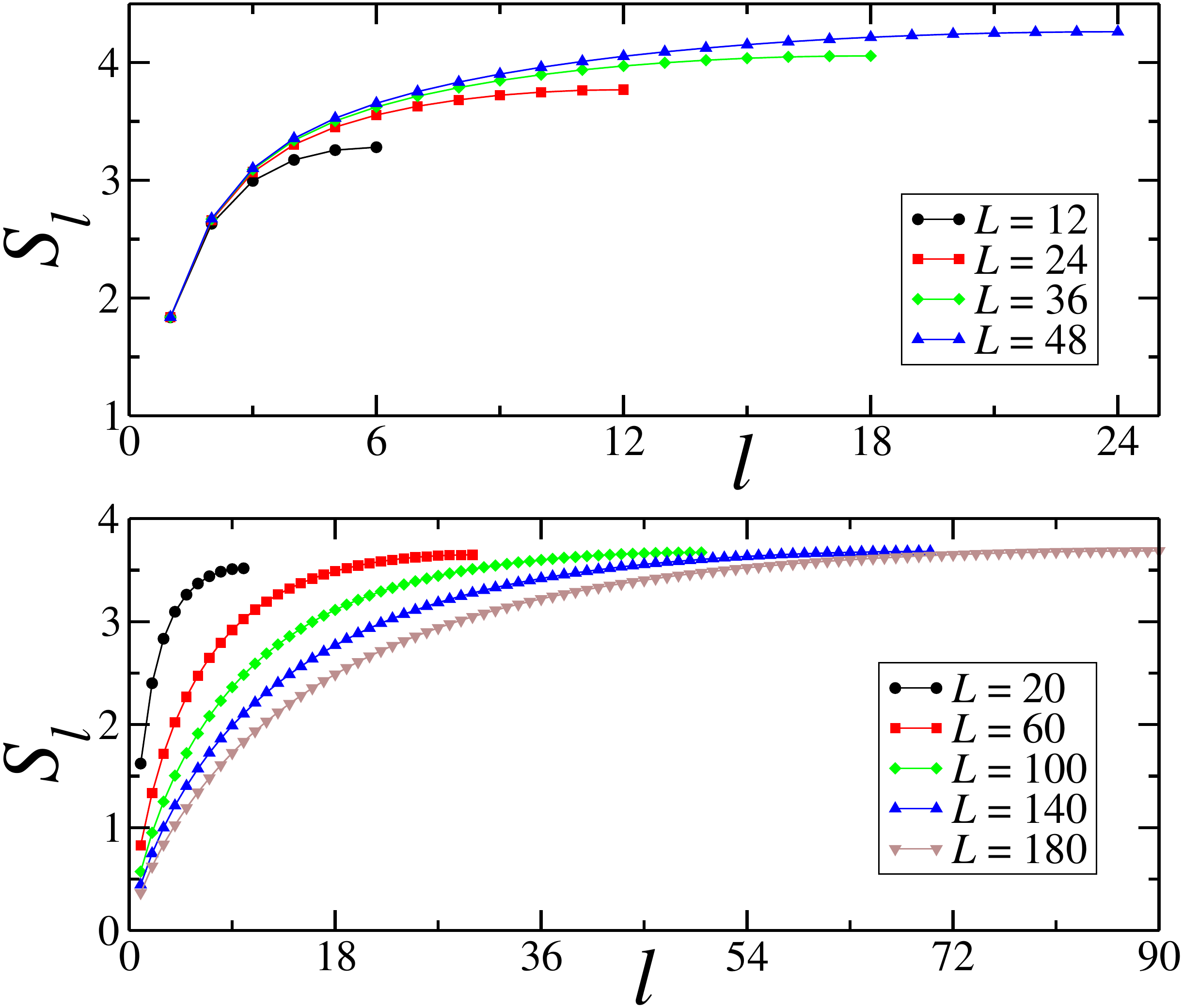}
\end{center}
\caption{Entanglement entropy $S_l$ for a block with $l$ sites, (\textit{upper panel}) considering a fixed filling $n=\frac{N}{2L} = \frac{1}{3}$, (\textit{lower panel}) or a fixed particle number sector $N=10$.}
\label{block.entropy}
\end{figure}

From the eigenvalues computed in this section, we can also compute the entanglement entropy of the block matrices, and see how it scales with 
the size of the block. We see in Fig.~\ref{block.entropy} a behavior typical of a gapless Hamiltonian, which does not scale as an area law.

\section{Edge modes}

As discussed in the main text, in order to directly characterize the localization length of the edge modes we compute the following matrix element:
$O_j = \bra{\psi_L(N)}_{oo} \hat a^\dagger_j \hat b_j \ket{\psi_L(N)}_{ee}$ (we consider here for simplicity the case of an even number of particles).
This task reduces to a particular counting of the suitable fermionic configurations, starting from the explicit expressions in Eq.~(4).

If we act with the $\hat V_j = \hat a^\dagger_j \hat b_j$ operator on $\ket{\psi_L(N)}_{ee}$, the
 only states $\ket{\vec{j}_{2n}}_a \otimes \ket{\vec{q}_{N-2n}}_a $ (see Eq.\eqref{eq:2wireWF} in the main text) which are
 not annnihilated bt $\hat V_j$ are those which have 
a particle at the $j$-th site of wire $b$ and a hole at the $j$-th site of wire $a$. 
After the action of $\hat V_j$, each of these configurations is changed by moving the fermion from the $j$-th site of wire $b$ to that of wire $a$.
Due to the anticommutation properties of fermionic operators, the state obtains the phase $(-1)^{(n_R^a + n_L^b)}$, where $n_R^a = \sum\limits_{r=j+1}^L n_r^a$ is the number of particles 
located at the right of the $j$-th site of the wire $a$, and $n_L^b = \sum\limits_{r=1}^{j-1} n_r^b$ the number 
of particles located at the left of the $j$-th site of the wire $b$. 
These phases describe the parity of the configuration on 
the segments $[j+1,L]$ for the wire $a$ and  $[1,j-1]$ for 
the wire $b$. Since the ground state is an equal superposition of all configurations distributing $N$ particles between the two wires
 and fixing the parity of the number of particles for each wire, $n_R^a$ varies from a minimum value equal to $\max(0,N-j)$ to
a maximum value $\min(N,L-j)$. Analogous relations exist for $n_R^b$. The matrix
 element $O_j$ is thus related to the simple counting of particle configurations
 taking into account the phase $(-1)^{(n_R^a + n_L^b)}$.
 
Let us discuss this in more detail.
The total number of fermionic configurations which originate the state $\ket{\psi_L(N)}_{ee}$ is 
\begin{equation}
\sum_n \underbrace{\binom{L}{2n}}_{b-wire} \underbrace{\binom{L}{N-2n}}_{a-wire},
\end{equation}
whereas those which have a particle at the $j$-th site of wire $b$ and a hole at the $j$-th site of wire $a$ are
\begin{equation}
\sum_n \underbrace{\binom{L-1}{2n-1}}_{b-wire} \underbrace{\binom{L-1}{N-2n}}_{a-wire} = \sum_n \underbrace{\left( \sum_{n_L^b}\binom{j-1}{n_L^b}\binom{L-j}{2n-1-n_L^b} \right)}_{b-wire} \,\, \underbrace{\left( \sum_{n_R^a}\binom{L-j}{n_R^a}\binom{j-1}{N-2n-n_R^a} \right)}_{a-wire}.
\end{equation}
Taking into account the phase $(-1)^{(n_R^a + n_L^b)}$, we have that
\begin{equation}
\bra{\psi_L(N)}_{oo} \hat a^\dagger_j \hat b_j \ket{\psi_L(N)}_{ee} = \frac{1}{\sqrt{\mathcal{N}_{ee,L,N} \mathcal{N}_{oo,L,N}}} \sum\limits_{n,n_L^b,n_R^a} (-1)^{(n_L^b+n_R^a)} \underbrace{\binom{j-1}{n_L^b}\binom{L-j}{2n-1-n_L^b} }_{b-wire} \,\, \underbrace{\binom{L-j}{n_R^a}\binom{j-1}{N-2n-n_R^a} }_{a-wire}
\label{exp.val.pert.local.jj}
\end{equation}

Following the same steps, it is not hard to see that for a more general number-conserving two-body operator we have:
\begin{equation}
\bra{\psi_L(N)}_{oo} \hat a^\dagger_j \hat b_{r} \ket{\psi_L(N)}_{ee} = \frac{1}{\sqrt{\mathcal{N}_{ee,L,N} \mathcal{N}_{oo,L,N}}} \sum\limits_{n,n_L^b,n_R^a} (-1)^{(n_L^b+n_R^a)} \underbrace{\binom{r-1}{n_L^b}\binom{L-r}{2n-1-n_L^b} }_{b-wire} \,\, \underbrace{\binom{L-j}{n_R^a}\binom{j-1}{N-2n-n_R^a} }_{a-wire}.
\end{equation}
for $(L+r)-j > 1$.
In particular,  edge-edge correlations have a particularly simple expression in the thermodynamic limit. Let us demonstrate that starting from:
\begin{eqnarray}\label{eq:edge:pert}
\bra{\psi_L(N)}_{oo} \hat a^\dagger_1 \hat b_{L} \ket{\psi_L(N)}_{ee} &=& \frac{1}{\sqrt{\mathcal{N}_{ee,L,N} \mathcal{N}_{oo,L,N}}} \sum\limits_{n} (-1)^{(N-1)} \underbrace{\binom{L-1}{2n-1}}_{b-wire} \,\, \underbrace{\binom{L-1}{N-2n }}_{a-wire};\\
\bra{\psi_L(N)}_{oo} \hat a^\dagger_L \hat b_{1} \ket{\psi_L(N)}_{ee} &=& \frac{1}{\sqrt{\mathcal{N}_{ee,L,N} \mathcal{N}_{oo,L,N}}} \sum\limits_{n} \underbrace{\binom{L-1}{2n-1}}_{b-wire} \,\, \underbrace{\binom{L-1}{N-2n }}_{a-wire}.
\end{eqnarray}
Using the Chu-Vandermonde identity, 
$\sum_{k=0}^r \binom{m}{k}\binom{n}{r-k} = \binom{m+n}{r}$, 
which holds for non-negative integer $m,n,r$,
 we obtain that in the limit of large lattices the following is true:
\begin{eqnarray}
\sum_n \underbrace{\binom{L-1}{2n-1}}_{b-wire}\underbrace{\binom{L-1}{N-2n-1}}_{a-wire} &\approx& \frac{1}{2}\binom{2L-2}{N-1};
\qquad 
\mathcal{N}_{ee(oo),L,N} \approx \frac{1}{2}\binom{2L}{N}.
\end{eqnarray}
Thus,  edge-edge correlations reduce to:
\begin{equation}
  \bra{\psi_L(N)}_{oo} \hat a^\dagger_1 \hat b_{L} \ket{\psi_L(N)}_{ee} \approx \frac{\nu (1-\nu) }{ (1-\frac{1}{2L})}
  \overset{L \to \infty}{\longrightarrow} \nu(1-\nu).
\end{equation}
Similar expressions exist for $\bra{\psi_L(N)}_{oo} \hat a^\dagger_L \hat b_{1} \ket{\psi_L(N)}_{ee}$. Note that, if $N$ is $odd$, a minus sign appears due to the overall phase $(-1)^{(N-1)}$ in Eq.~\eqref{eq:edge:pert}:
$\bra{\psi_L(N)}_{oo} \hat a^\dagger_1 \hat b_{L} \ket{\psi_L(N)}_{ee} \xrightarrow{L\rightarrow \infty} - \nu(1-\nu)$.

\section{Single-particle and superfluid correlations}

In a general way, any ground state observable can be computed as in the previous section through  a simple counting of suitable configurations. In this 
section we evaluate the single particle correlations $\langle \hat a^\dagger_j \hat a_{j+r} \rangle$, 
as well as the superfluid correlations $\langle \hat a^\dagger_i \hat a^\dagger_{i+1} \hat a_{j+1} \hat a_{j} \rangle$. 
We skip unnecessary details and focus mainly on the presentation of the final results.
We only consider the ground states for even values of $N$ because the odd case is mathematically equivalent.

\subsection{Single particle correlations:} 

\begin{align}
\bra{\psi_L(N)}_{ee(oo)}& \hat a^\dagger_j \hat a_{j+r} \ket{\psi_L(N)}_{ee(oo)} =\\
\nonumber \\
& \left\{ \begin{array}{ll} \frac{1}{\mathcal{N}_{ee(oo),L,N}} \sum\limits_{n,{n_{(j,r)}^a}} (-1)^{n_{(j,r)}^a} \underbrace{\binom{L}{2n (2n+1)} }_{b-wire} \underbrace{\binom{r-1}{{n_{(j,r)}^a}}\binom{L-r-1}{N-2n(2n+1)-1-{n_{(j,r)}^a}} }_{a-wire}, &\, \mbox{if } r>1; \\ \\
 \frac{1}{\mathcal{N}_{ee(oo),L,N}} \sum\limits_{n} \underbrace{\binom{L}{2n(2n+1)} }_{b-wire} \,\, \underbrace{\binom{L-r-1}{N-2n(2n+1)-1} }_{a-wire}, & \, \mbox{if } r\leq 1. \end{array} \right.
 \label{sp.exp.value.eveneven}
\end{align}
Here, $n_{(j,r)}^a = \sum_{i=1}^{r-1} n_{j+i}^a$ is the number of particles between the sites $j$ and $j+r$ of the wire $a$, which varies from a minimum of zero (where all the particles are in the wire $b$), to a maximum value equal to $\min(N-1,r-1)$ (where either all the remaining $N-1$ particles lie between these sites or the wire segment is completely filled).

\subsection{P-wave superfluid correlations:}
Pairing correlations characterize the model defined by Eq.~(3) and are of p-wave nature. Their behaviour is 
captured by the pairing operator $\hat a_{j+1} \hat a_j$. At the exactly solvable point, $\lambda =1$, these correlations can be 
computed analytically. The evaluation proceeds as in the previous cases, relying on the counting of the suitable fermionic configurations. The 
result for the p-wave correlation reads:
%
\begin{equation}
\bra{\psi_L(N)}_{ee(oo)} \hat a^\dagger_i \hat a^\dagger_{i+1} \hat a_{j+1} \hat a_j \ket{\psi_L(N)}_{ee(oo)}
= \frac{1}{\mathcal{N}_{ee(oo)}(L,N)}\sum_n \underbrace{\binom{L}{2n}}_{b-wire}\underbrace{\binom{L-4}{N-2n-2}}_{a-wire},
\qquad i+1<j.
\end{equation}
Using the Chu-Vandermonde identity, we obtain in the limit of large lattices,
\begin{eqnarray}
\sum_n \underbrace{\binom{L}{2n}}_{b-wire}\underbrace{\binom{L-4}{N-2n-2}}_{a-wire} &\approx& \frac{1}{2}\binom{2L-4}{N-2},
\end{eqnarray}
with $\mathcal{N}_{ee(oo)}(L,N) \approx \frac{1}{2}\binom{2L}{N}$.
At large distances the p-wave pairing correlations saturate to
\begin{equation}
  \bra{\psi_L(N)}_{ee(oo)} \hat a^\dagger_i \hat a^\dagger_{i+1} \hat a_{j+1} \hat a_j \ket{\psi_L(N)}_{ee(oo)} 
\approx \frac{2^4 \nu^2 (1-\nu)^2 L^4 + \mathcal{O}(L^3)}{2^4 L^4 + \mathcal{O}(L^3)} \overset{L \to \infty}{\longrightarrow} \nu^2(1-\nu)^2.
\end{equation}
This non-zero value for large $|i-j|$ distances hints to the fact that there are no sound-like modes in the exactly-solvable model, since they would lead to 
the algebraic decay of the p-wave correlations.

\section{Braiding}

In this section we  demonstrate  the properties of 
the braiding operator $\hat{B}_{aR,bR}(j) = (\hat{\mathcal{I}} + \hat{Z}_{aR,bR,j})/\sqrt{2}$ presented in the main text. 
To that end, it is useful to first study the action of  $\hat{Z}_{aR,bR,j}$, starting from the simpler case $j=1$:
\begin{equation}
\hat Z_{aR,bR,1} = \frac{\hat X_{aR,bR,1}}{\mathcal{F}_1};
\qquad \hat X_{aR,bR,1} = \ad{L}b_{L} - \bd{L}a_{L} ;
\qquad \mathcal F_1 = \sqrt{1 - [\nu^2 + (1-\nu)^2]};
\qquad \nu = \frac{N}{2L}.
\end{equation}
The action of $\ad{L}b_{L}$ on $ \ket{\psi_L(N)}_{ee}$ produces an unnormalized state proportional to the equal-weighted superposition
of all fermionic configurations with (i) $N$ fermions, (ii) odd wire parities, (iii) a hole in the $L$-th site of the wire $a$, and (iv) a particle in the $L$-th site of the wire $b$. 
Due to the anticommuting properties of fermionic operators,
this state gets a global phase $(-1)^{(N-1)+(N_a)} = (-1)^{N_b-1}$ (recall that $(-1)^{N_a}$ and $(-1)^{N_b}$ are well-defined although $N_a$ and $N_b$ are not fixed).
With similar reasoning one can also characterize $\bd{L}a_{L} \ket{\psi_L(N)}_{ee}$ (here the hole (fermion) is in the $L$-th site of the wire $a$ ($b$) and the global phase is $(-1)^{(N_a-1)+(N-1)} = (-1)^{N_b}$). 
The state $\hat{Z}_{aR,bR,1} \ket{\psi_L(N)}_{ee}$ is the normalized state described by the equal weighted superposition of all fermionic configurations with (i) $N$ fermions and (ii) odd wire parities and without (iii) the simultaneous presence of two holes or two particles at the $L$-th site of both wires $a$ and $b$. 
The normalization constant $F_1$  corresponds to the ``fidelity'' of the state with the ground state with such local parities, 
   $\mathcal{F}_1 = \sqrt{\bra{\psi_L(N)}_{oo} \hat{Z}_{aR,bR,1} \ket{\psi_L(N)}_{ee}}
= \sqrt{1 - [\nu^{2} + (1-\nu)^{2}]}$.

The operator $\hat{Z}_{aR,bR,j}$ acts on the last $j$ sites of the wires. Let us consider for example $j=2$:
\begin{align}
 \hat Z_{aR,bR,2} =& \frac{\hat X_{aR,bR,1} + \hat Y_{aR,bR,1}\hat X_{aR,bR,2}}{\mathcal{F}_2};
 \nonumber \\
 \hat Y_{aR,bR,1}\hat X_{aR,bR,2}
 =&  \left(n_{L}^a n_{L}^b + (1-n_{L}^a)(1-n_{L}^b) \right)
 \left( \ad{L-1}b_{L-1} - \bd{L-1}a_{L-1} \right);
\qquad
\mathcal F_2 = \sqrt{1 - [\nu^4 + (1-\nu)^4]}.
\end{align}
Note that the term $\hat Y_{aR,bR,1}\hat X_{aR,bR,2} \ket{\psi_L(N)}_{ee}$ acts only on the fermionic configurations which were missing in $\hat{X}_{aR,bR,1} \ket{\psi_L(N)}_{ee}$. However, it is clear that $\hat Z_{aR,bR,2}\ket{\psi_L(N)}_{ee}$ does not contain any configuration with four holes or four particles in the $L$-th and $(L-1)$-th sites of both wires. Whereas this still makes $\hat Z_{aR,bR,2}\ket{\psi_L(N)}_{ee}$ different from $\ket{\psi_L(N)}_{oo}$, it is a considerable improvement with respect to the previous case. In general, the state $\hat{Z}_{aR,bR,j} \ket{\psi_L(N)}_{e e}$ is the equal weighted superposition of all fermionic configurations with (i) $N$ fermions and (ii) odd wire parities, and without (iii) the simultaneous presence of $2j$ holes or $2j$ particles in the last $j$ sites of both wires. $\mathcal{F}_j$ corresponds to the fidelity of the state
   to the ground state with such local parity pattern, 
   $\mathcal{F}_j = \sqrt{\bra{\psi_L(N)}_{oo} \hat{Z}_{aR,bR,j} \ket{\psi_L(N)}_{ee}}
= \sqrt{1 - \left[\nu^{2j} + (1-\nu)^{2j}\right]}$.
 For a large enough $j < L/2$,
 the difference between $\hat Z_{aR,bR,j}\ket{\psi_L(N)}_{ee}$ and $\ket{\psi_L(N)}_{oo}$
 becomes exponentially small, so that the above operator 
 \textit{in the zero-energy subspace} reads
 \begin{align}
 \hat Z_{aR,bR,j} \sim & \sum_{N \, even}\left(\ket{\psi_L(N)}_{ee}\bra{\psi_L(N)}_{oo} -
  \ket{\psi_L(N)}_{oo} \bra{\psi_L(N)}_{ee}\right) + \nonumber \\
 &  + \sum_{N\, odd}  \left(\ket{\psi_L(N)}_{oe}\bra{\psi_L(N)}_{eo} -
  \ket{\psi_L(N)}_{eo} \bra{\psi_L(N)}_{oe}\right).
  \label{eq:Z:op}
 \end{align}
For $j$ and $L$ are such that the error is negligible, we define the \textit{number conserving} operator
  $\hat{R}_{aR,bR}  = \hat{B}_{aR,bR}(j)$. 

We observe that $\hat{R}_{aR,bR}$ acts on the zero-energy states in the same way done in the conventional \textit{number non-conserving} scenario of two neighboring Kitaev wires by the braiding operator $\hat R'_{aR,bR} = e^{\frac{\pi}{4} \gamma_{aR} \gamma_{bR}} = (\mathcal{I} + \gamma_{aR} \gamma_{bR})/\sqrt{2}$, where $\gamma_{m\Lambda}$ are the zero-energy Majorana  operators exponentially localized at the $\Lambda =R,L$ edge of the wire $m=a,b$.
In order to verify this explicitly, we first recall that 
the number non-conserving edges Majoranas are related to a non-local fermion as
$\hat f_{m} = \hat \gamma_{m L} - i\gamma_{mR}$  which is the Bogoliubov zero-energy mode. 
The two degenerate ground states of the wire $m=a,b$, $\ket{\psi}_{m\sigma}$ ($\sigma$ is the parity of  the number of fermions, even, $e$ or odd, $o$, and labels the two ground states), correspond to the presence or absence of the non-local fermion $\hat f_m$: $\hat f_m \ket{\psi}_{me} = 0$, and $\hat f^{\dagger}_m \ket{\psi}_{me} = \ket{\psi}_{mo}$. 
Using the inverse relations $\hat \gamma_{mL} \propto \left(\hat f_{m} + \hat f_{m}^\dagger \right)$ and $\hat \gamma_{mR} \propto i \left(\hat f_{m} - \hat f_{m}^\dagger \right)$, it is now easy to see that:   
\begin{equation}
  \hat \gamma_{aR} \hat \gamma_{bR} \ket{\psi}_{a \sigma} \ket{\psi}_{b \tau} = p_\sigma (\hat \gamma_{aR} \ket{\psi}_{a \sigma}) (\hat \gamma_{bR} \ket{\psi}_{b \tau}) 
   = p_\sigma  (-i p_\sigma) (-i p_\tau) \ket{\psi}_{a \bar \sigma} \ket{\psi}_{b \bar \tau} 
   = -p_\tau \ket{\psi}_{a \bar \sigma} \ket{\psi}_{b \bar \tau}
  \end{equation}
  where $p_\sigma = 1$ for $\sigma=e$ and $p_\sigma = -1$ for $\sigma=o$;  $\bar{\sigma}, \bar{\tau}$ are the flipped $\sigma, \tau$.
As claimed, $\hat \gamma_{aR} \hat \gamma_{bR}$ acts on the ground space in a way which is completely analogous to that of $\hat{Z}_{ab}^R$ in Eq.~\eqref{eq:Z:op} for the number-conserving model. The equivalence of $\hat R_{aR,bR}$ and $\hat R'_{aR,bR}$ follows directly.
  
The unitarity of the braiding operator $\hat R_{aR,bR}$ restricted to the ground subspace can be proved explicitly. Let us first notice that:
\begin{align}
\bra{\psi_L(N)}_{\sigma\tau} \hat{R}_{aR,bR}^{\dagger} \hat{R}_{aR,bR} \ket{\psi_L(N)}_{\sigma'\tau'}
=& \frac 12 \bra{\psi_L(N)}_{\sigma\tau} \left( \mathcal{I} + \hat{Z}_{aR,bR,j} + \hat{Z}_{aR,bR,j}^{\dagger}   +  \hat{Z}_{aR,bR,j}^{\dagger} \hat{Z}_{aR,bR,j} \right) \ket{\psi_L(N)}_{\sigma'\tau'} \nonumber \\
=& \frac 12 \bra{\psi_L(N)}_{\sigma\tau} \left( \mathcal{I} +  \hat{Z}_{aR,bR,j}^{\dagger} \hat{Z}_{aR,bR,j} \right) \ket{\psi_L(N)}_{\sigma'\tau'} 
= \delta_{\sigma,\sigma'} \delta_{\tau,\tau'}
\end{align}
where in the second line we used the fact that $\hat{Z}_{aR,bR,j}$ is anti-Hermitian. 
Thus, in the ground state subspace we have,
\begin{equation}
\hat{P}_g \, \hat{R}_{aR,bR}^{\dagger} \, \hat{R}_{aR,bR} \, \hat{P}_g^{\dagger} = \hat{P}_g
\end{equation}

With similar procedures we can define other braiding operators with completely analogous properties. For example, let us consider the transformation $i \leftrightarrow (L+1-i)$ which maps an operator at site $i$ to site $(L+1-i)$, thus mapping the right edge to the left one and viceversa. When we apply it to $\hat Z_{aR,bR,j}$, we can define $\hat Z_{aL,bL,j}$, which is exponentially localized at the left edge of the wire. For $j$ and $L$ large enough, its explicit expression is:
\begin{align}
\hat Z_{aL,bL,j} \sim & \sum_{N \, even}\left(- \ket{\psi_L(N)}_{ee}\bra{\psi_L(N)}_{oo} + \ket{\psi_L(N)}_{oo} \bra{\psi_L(N)}_{ee}\right)  \nonumber \\
&   + \sum_{N\, odd} \left(\ket{\psi_L(N)}_{oe}\bra{\psi_L(N)}_{eo} -
 \ket{\psi_L(N)}_{eo} \bra{\psi_L(N)}_{oe}\right).
\end{align}
The braiding operator $\hat R_{aL,bL}$ can be defined, and an explicit calculation shows that it is unitary and that it resembles the operator $\hat R'_{aL,bL}$ for the number non-conserving case.

As a second example, let us consider the transformation $b_{L+1-j} \to - i a_{j}$ for $j<L/2$ which maps the right edge of the wire $b$ to the left edge of the wire $a$, leaving the other two edges unchanged. When we apply it to $\hat Z_{aR,bR,j}$, we can define the operator $\hat Z_{aR,aL,j}$, whose explicit expression for $j$ and $L$ large enough is:
\begin{align}
\hat Z_{aR,aL,j} \sim & \sum_{N \, even} i\, \left( \ket{\psi_L(N)}_{ee}\bra{\psi_L(N)}_{ee} -
 \ket{\psi_L(N)}_{oo} \bra{\psi_L(N)}_{oo}\right)  \nonumber \\
&   + \sum_{N\, odd}  i \, \left(\ket{\psi_L(N)}_{eo}\bra{\psi_L(N)}_{eo} -
 \ket{\psi_L(N)}_{oe} \bra{\psi_L(N)}_{oe}\right).
\end{align}
Again, everything follows as before.
In general, with this method one can define sixteen braiding operators $\hat R_{m\Lambda,m'\Lambda}$, with $m,m'=a,b$ labeling the wires and $\Lambda,\Lambda'=L,R$ labeling the left and right edge.

Let us conclude with an explicit verification of the non-abelian character of these operators; to this aim, we initialize the system in the state $\ket{\psi_L(N)}_{ee} $. We then perform two braiding operations on the edges in different sequences:
 \begin{eqnarray}
\left(\hat{R}_{aR,aL} \hat{R}_{aR,bR} - \hat{R}_{aR,bR} \hat{R}_{aR,aL} \right) \ket{\psi_L(N)}_{ee} = 
\frac 12 \left(\hat{Z}_{aR,aL,j} \hat{Z}_{aR,bR,j} -  \hat{Z}_{aR,bR,j} \hat{Z}_{aR,aL,j} \right) \ket{\psi_L(N)}_{ee}  
=   i \ket{\psi_L(N)}_{oo}.\nonumber\\
 \end{eqnarray}
The result shows explicitly the non-commutativity of the braiding operations.  


\end{document}